\documentclass[conference]{IEEEtran}
\IEEEoverridecommandlockouts
\usepackage{cite}
\usepackage{amsmath,amssymb,amsfonts}
\usepackage{algorithmic}
\usepackage{graphicx}
\usepackage{textcomp}
\usepackage{xcolor}
\usepackage{amsthm}
\usepackage{overpic}
\usepackage{bm}
\usepackage{overpic}
\usepackage{float}
\usepackage{amsthm}
\usepackage{algorithm,algorithmic}

\theoremstyle{plain}
\newtheorem{theorem}{Theorem}

\newtheorem{lemma}{Lemma}
\newtheorem{corollary}{Corollary}

\makeatletter
\newcommand\fs@ruled@notop{\def\@fs@cfont{\bfseries}\let\@fs@capt\floatc@ruled
  \def\@fs@pre{}%
  \def\@fs@post{\kern2pt\hrule\relax}%
  \def\@fs@mid{\kern2pt\hrule\kern2pt}%
  \let\@fs@iftopcapt\iftrue}
\renewcommand\fst@algorithm{\fs@ruled@notop}
\makeatother

\begin{document}

\title{Fundamentals of Energy-Efficient Wireless Links: Optimal Ratios and Scaling Behaviors
}
\author{\thanks{This work was supported by the FFL18-0277 grant from the Swedish Foundation for Strategic Research.}

\IEEEauthorblockN{Anders Enqvist$^*$, {\"O}zlem Tu\u{g}fe Demir$^\dagger$, Cicek Cavdar$^*$,  Emil Bj{\"o}rnson$^*$}
\IEEEauthorblockA{ {$^*$Department of Computer Science, KTH Royal Institute of Technology, Kista, Sweden}
\\ {$^\dagger$Department of Electrical-Electronics Engineering, TOBB University of Economics and Technology, Ankara, Türkiye
		} \\
		{Email: enqv@kth.se, ozlemtugfedemir@etu.edu.tr,  cavdar@kth.se, emilbjo@kth.se }
}

}


\maketitle

\begin{abstract}
In this paper, we examine the energy efficiency (EE) of a base station (BS) with multiple antennas. We use a state-of-the-art power consumption model, taking into account the passive and active parts of the transceiver circuitry, including the effects of radiated power, signal processing, and passive consumption. The paper treats the transmit power, bandwidth, and number of antennas as the optimization variables. We provide novel closed-form solutions for the optimal ratios of power per unit bandwidth and power per transmit antenna. We present a novel algorithm that jointly optimizes these variables to achieve maximum EE, while fulfilling constraints on the variable ranges. We also discover a new relationship between the radiated power and the passive transceiver power consumption. We provide analytical insight into whether using maximum power or bandwidth is optimal {\color{black} and how many antennas a BS should utilize}.
\end{abstract}


\section{Introduction}


The exponential growth in data rates within wireless communication systems, as dictated by Cooper's law, has substantially raised energy consumption. Anticipating continued exponential growth in traffic demands~\cite{ericssonmobilityreport2023}, it is imperative to enhance the energy efficiency (EE) of wireless communication technologies. It is often defined as the data rate divided by the related power consumption. By unraveling the fundamental behaviors and limitations of EE in wireless communication systems, we can uncover innovative approaches and technologies that promise more sustainable and efficient wireless networks in the future. While theoretical tools for EE optimization have been developed for decades \cite{Verdu1990a,ZapponeNowPublishers2015}, the willingness to make improvements in practice has garnered heightened attention in recent times, as evidenced by ITU targets \cite{series2017minimum}. Much of the technology development has focused on increasing data rates through the introduction of massive MIMO (multiple-input multiple-output) \cite{bjornson2017massive} and increasing bandwidth in mmWave and terahertz bands \cite{Rappaport2019a}.
Additionally, emerging technologies such as Reconfigurable Intelligent Surfaces (RIS) \cite{RIS-EE} promise reductions in energy consumption in future networks. A recent survey on how these and other techniques can lead to power consumption reductions can be found in \cite{lopez2022survey}.


\subsection{Prior Work and Motivations}

EE optimization was pioneered in \cite{Verdu1990a}, which emphasized a strong tradeoff between EE and rates.
 The paper \cite{auer2011much} is an early work on how much power is needed to run a wireless communication system, which is much more than just transmit power. The papers \cite{Bjornson2016aabb,Zappone2023tradeoff} present a network model that considers the optimization of the area power consumption (PC) and area EE under rate constraints. 
 A PC model with parameters that capture the fundamental behaviors of base stations (BSs) was presented in \cite{bjornson2015a}. It has since been extended to include PC due to carrier aggregation in \cite{lopez2021energy}. 
 In \cite{piovesan2022machine,piovesan2022power}, machine learning (ML) was employed for curve fitting real-world data to adapt a PC model capturing essential aspects. These works showcase the power of ML in wireless communication modeling and demonstrate that an analytical linear model, with appropriately chosen constants, can accurately represent the power consumption in current BSs. 

While prior works have examined different EE optimization problems in various complex wireless networks, this paper returns to the fundamentals: we consider a single communication link between a multi-antenna BS and a single-antenna user equipment (UE).
When one truly seeks the EE-optimal communication system design, the PC model has a huge impact on the solution. If one only accounts for the transmit power, the optimum is achieved as the rate approaches zero \cite{Verdu1990a}. By contrast, \cite{bjornson2018energy} studied the upper EE limit with a more detailed PC model and demonstrated that very different operating points might be approached in the distant future.

 In our new analysis, we uncover an intriguing new relationship between transmit power and transceiver fixed power at EE-optimal solution, which contributes to the fundamental understanding of how to build energy-efficient wireless systems. We discover new analytical relations between the transmit power, bandwidth, and number of antennas, and how these are related to the hardware characteristics. We also develop an algorithm for jointly optimizing these parameters.

\subsection{Contributions}

In this paper we aim to answer the research questions:

\begin{itemize}
    \item How should the transmit power, bandwidth, and number of antennas be jointly configured to maximize EE?
    \item Are there any tangible relationships between these parameters at the optimal solution?
\end{itemize}

One significant departure from previous papers is our emphasis on analytical scaling behaviors. While prior research primarily focused on optimizing individual parameters, our work extends these models to explore the global optimum.



\section{System Model}


We analyze and optimize the energy efficiency of the link between a BS using $M$ antennas and a single-antenna UE. The carrier bandwidth is $B$ and the channel is represented by $\mathbf{h}\in \mathbb{C}^M$, where the squared magnitude of each entry is $\beta$. Hence, $\|\mathbf{h}\|^2=M\beta$. 
This is a typical model for a line-of-sight channel. The received downlink signal $y$ is given by
\begin{equation}
    y=\mathbf{h}^{T}\mathbf{p}x+n,
\end{equation}
where $x$ is the data signal, $n\sim \mathcal{N}_\mathbb{C}(0,B N_0 )$ is the independent receiver noise, and $\mathbf{p}\in\mathbb{C}^{M}$ is the unit-norm precoding vector.  Assuming the BS has perfect channel state information (CSI), the capacity of this multiple-input-single-output (MISO) channel is achieved by $x \sim \mathcal{N}_\mathbb{C}(0,P)$, where $P$ is the transmit power budget, and the precoding vector $\mathbf{p}=\mathbf{h}^*/\|\mathbf{h}\|$.  The data rate given by the capacity is \cite{Lo1999a,bjornson2017massive}

\begin{equation}
\label{eq:capacity}
    C=B \log_2 \left( 1 + \frac{M P \beta}{B N_0} \right).
\end{equation}

Notice that we have made modeling assumptions that enable exact mathematical analysis. {\color{black} However, the qualitative insights also hold for other types of channel realizations such as Rayleigh fading with a variance of $\beta$.}

\subsection{Power Consumption}

The power consumption (PC) is modeled as in \cite{lopez2021energy}, for single-layer transmission in a single band to a single-antenna UE. The total PC at the BS is
\begin{align}
\label{PC_model}\mathrm{PC}=&P/\kappa+P_\mathrm{FIX}+P_\mathrm{SYN}+D_0 M + D_1 M+\eta C, 
\end{align}
where $\kappa \in (0,1]$ is the power amplifier (PA) efficiency and $P_\mathrm{FIX}$ is the load-independent power consumption required for cooling, control signaling, backhaul infrastructure, and baseband processors. $P_\mathrm{SYN}$ is the load-independent power consumed by the local oscillator. $D_0$ is the power consumed by each transceiver chain (antenna port) of the BS (e.g., converters, mixer, filters, etc.). $D_1$ is the power consumed by the signal processing at the BS that scales with the number of antennas, including channel estimation and precoding. $\eta$ regulates the power consumed by the signal coding at the BS and 
the backhaul signaling, both of which is proportional to the capacity $C$. To simplify the notation and expose the optimization variables, we rewrite \eqref{PC_model} as 



\begin{equation}
\label{eq:PC}
\mathrm{PC}=P/\kappa + \mu+( D_0+\nu B)  M + \eta B \log_2 \left( 1 + \frac{M P \beta}{B N_0} \right),
\end{equation}
where $\mu=P_\mathrm{FIX}+P_\mathrm{SYN}$ denotes the fixed circuit power consumption from circuitry and synchronization and $\nu=D_1/B$ is introduced to highlight that the signal processing is carried out on the sampling rate (which is proportional to the bandwidth).

%


\subsection{Energy Efficiency}

In this paper, we focus our efforts on optimizing energy efficiency (EE) \cite{Verdu1990a,bjornson2017massive}, defined as the amount of data transferred per unit energy (measured in bit/Joule and equivalently bit/s/Watt). Dividing the channel capacity in \eqref{eq:capacity} by the power consumption in \eqref{eq:PC}, we can define the $\mathrm{EE}$ as

\begin{equation} \label{eq:EE}
\mathrm{EE} =  \frac{B \log_2 \left( 1 + \frac{M P \beta}{B N_0} \right) }{P/\kappa + \mu  + ( D_0+\nu B )M + \eta B \log_2 \left( 1 + \frac{M P \beta}{B N_0}  \right) } 
\end{equation}

In the following sections, we will study the scaling behaviors of the EE with the bandwidth $B$, power $P$, and number of antennas $M$. In particular, we will derive the optimal pairwise ratios of these three design parameters, and then design an algorithm that finds the global optimum.

\section{EE-Optimal Parameter Ratios}

In this section, we will prove that the optimization variables $B$, $P$, and $M$ tend to satisfy specific ratios at the EE-optimal system operation. These results serve as design guidelines.

\subsection{Power per Bandwidth}
By dividing the numerator and denominator of \eqref{eq:EE} by $B$, the EE can be rewritten as

\begin{equation} \label{eq:EEdivB}
\mathrm{EE} \!=\!  \frac{\log_2 \left( 1 + \frac{M P \beta}{B N_0} \right) }{P/(\kappa B) + \mu/B  + D_0 M/B+\nu M + \eta \log_2 \left( 1 \!+\! \frac{M P \beta}{B N_0}  \right) }.
\end{equation}
It is apparent from \eqref{eq:EEdivB} that power and bandwidth mostly appear as a ratio $P/B$, which is the power spectral density. The terms $D_0 M /B$ and $\mu/B$ are the only ones that do not fit this structure. However, in practical situations in which enough bandwidth is available, these terms are expected to be negligible compared to the terms that depend on both the bandwidth and power. This leads to the following result:
\begin{theorem} \label{maximum-EE-lemma}
When the term $\mu/B  + D_0 M/B$ is negligible, the $\mathrm{EE}$ in \eqref{eq:EEdivB} is maximized when $P$ and $B$ satisfy the ratio 
\begin{equation} \label{eq:optimal-ratio}
\frac{P}{B} = N_0 \frac{e^{u}-1}{M \beta},
\end{equation}
where
\begin{equation}
u= W \left(\frac{\kappa M^2 \beta  \nu}{N_0 e} -\frac{1}{e}\right)+1
\end{equation}
and $W (\cdot)$ denotes the Lambert W function, defined by the equation $x = W(x)e^{W(x)}$ for any $x \in \mathbb{C}$.
\end{theorem}
\begin{IEEEproof}
By defining $z= P/B$, and letting $\mu,D_0 \to 0$, \eqref{eq:EEdivB} can be expressed as
\begin{equation} \label{eq:EE-first-step-SISO-circuit-power2-z}
  \frac{ \log_2 \left( 1 + \frac{M \beta}{N_0} z \right) }{\frac{z}{\kappa} + \nu M + \eta \log_2 \left( 1 + \frac{M \beta}{ N_0} z \right) }.
\end{equation}
The maximum in \eqref{eq:optimal-ratio} is obtained by utilizing \cite[Lem.~3]{bjornson2015a}.
\end{IEEEproof}




By inserting \eqref{eq:optimal-ratio} into \eqref{eq:EE-first-step-SISO-circuit-power2-z}, an upper bound on the maximum EE is obtained as a function of $M$ as
\begin{equation} \label{eq:EE_max}
    \mathrm{EE}_\mathrm{{max}}(M) = \frac{u \log_2(e)}{N_0\frac{e^u-1}{\kappa M \beta}+\nu M + \eta u \log_2(e)} ,
\end{equation}
where the effects of $\mu$ and $D_0$ have been neglected.

\begin{figure}[t!]
	\centering \vspace{-2mm}
	\includegraphics[width=\columnwidth]{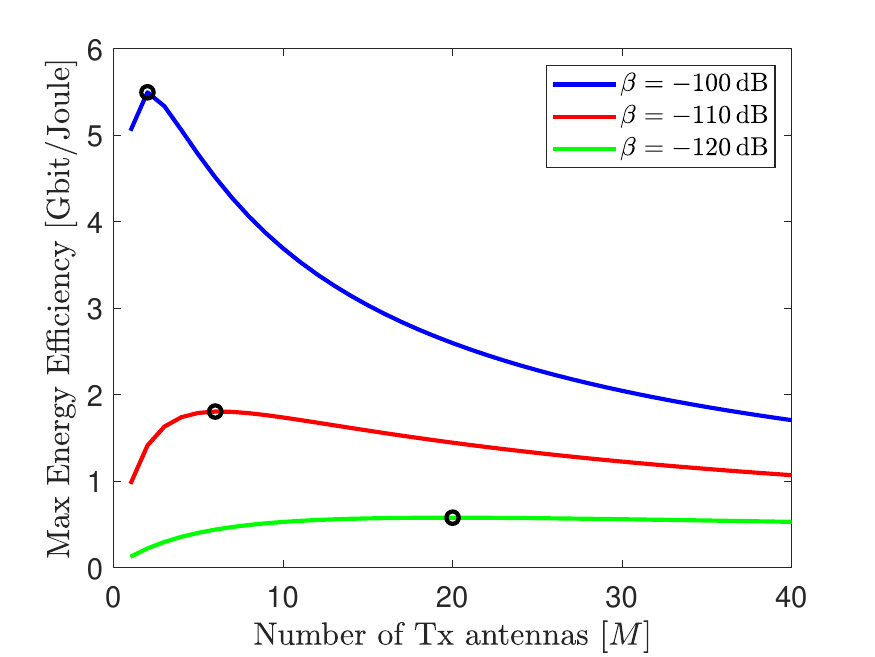}
	\caption{We have that the maximum energy efficiency $\mathrm{EE}_\mathrm{max}$ as defined in \eqref{eq:EE_max}  is attained for a finite number $M_\mathrm{opt}$ at an optimal ratio $P/B$. }
	\label{fig:MaxEEofM} \vspace{-2mm}
\end{figure}

We have $\mathrm{EE}_\mathrm{{max}}(M)>0$ for $M>0$ and $\lim_{M\to 0}\mathrm{EE}_\mathrm{{max}}(M)=\lim_{M\to \infty}\mathrm{EE}_\mathrm{{max}} =0$.  This means that there exists an optimal finite value $M_\mathrm{opt}$ that maximizes \eqref{eq:EE_max}. This is illustrated in Fig.~\ref{fig:MaxEEofM} with the simulation parameters being given in Table \ref{tab:table1}. We can see that the optimum EE (encircled) is obtained at $M=2$ for $\beta=-100\,$dB, $M=6$ for $\beta=-110\,$dB, and $M=20$ for $\beta=-120\,$dB. This indicates that the optimal $M$ grows as $\beta$ becomes smaller. For our set of constants, $M$ was increased by a factor $10$ when $\beta$ decreased $-20\,$dB. We obtain the optimal power spectral density: $P/B=19\,$mW/GHz for $\beta=-100\,$dB $P/B=80$\,mW/GHz for $\beta=-110\,$dB and $P/B=251$\,mW/GHz for $\beta=-120\,$dB. This means that the ratio $P/B$ should be shifted towards a higher value to overcome a larger pathloss.  {\color{black} Through optimizing $M$ we learn how to operate the BS to reach the optimum EE and how to design it with enough antennas depending on the pathloss in its environment.}

The resulting signal-to-noise ratio (SNR) is defined as 
\begin{equation}
    \mathrm{SNR}=\frac{M P \beta}{B N_0}
\end{equation}
and can also be inferred from Theorem \ref{maximum-EE-lemma}. In this example, the SNR for the optimal solution is $\mathrm{SNR}=6.00\,$dB if $\beta=-100\,$dB, $\mathrm{SNR}=5.71\,$dB if $\beta=-110\,$dB, and $\mathrm{SNR}=6.00\,$dB if $\beta=-120\,$dB. This means that 4-QAM is roughly the optimal modulation scheme. The capacity expression is slightly non-linear at this operating point, even if $P$ and $M$ enter linearly into the PC model.

The $\mathrm{EE}$ in \eqref{eq:EE_max} is achieved by any values of $P$ and $B$ with the ratio in \eqref{eq:optimal-ratio} and that are sufficiently large to make the term $\mu/B  + D_0 M/B$ negligible. Hence, we have the freedom to choose $B$ to achieve any desired data rate
\begin{equation}
C = B u \log_2(e).
\end{equation}
In other words, if $P$ and $B$ are not limited by external factors, the EE and rate are permitted to grow together---there is no tradeoff between them as conventionally claimed \cite{Verdu1990a,bjornson2017massive}.

\begin{table}[t!]
  \begin{center}
    \caption{Simulation Parameters}
    \label{tab:table1}
    \begin{tabular}{|l|r|} 
     \hline
      \textbf{Parameter} & \textbf{Value} \\
      \hline
      Passive circuit power: $\mu$ & $100$\,mW \\
      Transceiver chain power consumption: $D_0$ & $20$\,mW \\
      Sample processing power consumption: $\nu$ & $10^{-10}$\,J/sample \\
      Power amplifier efficiency: $\kappa$ & 0.4 \\
      Computational efficiency: $\eta$ & $10^{-11}$  J/bit\\
      Maximum bandwidth: $B_\mathrm{max}$ & $10$\,GHz \\
      Maximum power: $P_\mathrm{max}$ & 40\,dBm \\
      Maximum number of transmit antennas: $M_\mathrm{max}$ & $512$ \\
      Receiver noise power spectral density: $N_0$ & $-174$\,dBm/Hz \\
      Channel gain: $\beta$ & $-110$\,dB\\
       \hline
    \end{tabular}
  \end{center}
\end{table}

\subsection{Power per Antenna}
It is also possible to analytically optimize the transmit power per antenna $P/M$, which leads to further insights. In practice, there might be upper bounds on both of these parameters which prevent us from reaching the optimal ratio. However, in case $P$ and $M$ are not upper bounded, or the maximum of the EE in \eqref{eq:EE} is reached without invalidating the bound $M_\mathrm{max}$ and $P_\mathrm{max}$, the following is true:

\begin{theorem} \label{PoverM}
If the solution $(P_\mathrm{opt},M_\mathrm{opt})$ that maximizes the EE in \eqref{eq:EE} for a given value of $B$ satisfies $P_\mathrm{opt} \leq P_\mathrm{max}$ and $M_\mathrm{opt} \leq M_\mathrm{max}$, then the following relation holds:
\begin{equation} \label{eq:optimal-ratio-PM}
\frac{P_\mathrm{opt}}{M_\mathrm{opt}}=\kappa(D_0+\nu B).
\end{equation}

\end{theorem}

\begin{IEEEproof}
We assume $B$ is fixed and define $a=\beta/(BN_0)$, $b=1/\kappa$, $c = D_0+\nu B$. Then, the EE optimization problem with respect to $P$ and $M$ becomes
\begin{align} 
&\underset{P, M}{\textrm{minimize}}  \quad \frac{\mu  + bP+ cM}{B \log_2 \left( 1 + a M P \right) }.
\end{align}

We take the first-order derivatives of the objective function with respect to $P$ and $M$, and equate them to zero, which leads to 
\begin{align}
&\dfrac{\ln\left(2\right)\,b}{B\ln\left(aMP+1\right)} =\dfrac{\ln\left(2\right)\,aM\cdot\left(\mu+bP+cM\right)}{B\cdot\left(aMP+1\right)\ln^2\left(aMP+1\right)}, \\
& \dfrac{\ln\left(2\right)\,c}{B\ln\left(aMP+1\right)} =\dfrac{\ln\left(2\right)\,aP\cdot\left({\mu}+bP+cM\right)}{B\cdot\left(aMP+1\right)\ln^2\left(aMP+1\right)} .
\end{align}
If we divide the second equation by the first equation, we obtain \eqref{eq:optimal-ratio-PM}.
\end{IEEEproof}

Theorem \ref{PoverM} has several interesting implications. The right-hand-side in \eqref{eq:optimal-ratio-PM} grows as either $\kappa$, $D_0$, $\nu$ or $B$ increase. It is evident that as the computational cost $\nu B$ associated with an increased bandwidth grows or the PA is of higher quality (i.e., larger $\kappa$), we can afford to transmit more power per antenna when reaching the EE-optimal solution. Moreover, if we can afford to use more antennas to gain higher EE, we should also increase the transmit power to maintain the power per antenna.

Further insights are obtained by rearranging \eqref{eq:optimal-ratio-PM} so that
\begin{equation} \label{eq:optimal-ratio-PM2}
    \frac{P_\mathrm{opt}}{\kappa}=(D_0+\nu B)M_\mathrm{opt}.
\end{equation}
We recognize that $P/\kappa + (D_0+\nu B)M$ appear directly in the PC consumption model in \eqref{eq:PC}. Hence, \eqref{eq:optimal-ratio-PM2} tells us that at the EE-optimal point, the input transmit power $P/\kappa$ is always identical to the power $(D_0+\nu B)M$, i.e., the passive power consumption in the transceiver chains for all the antennas $D_0 M$ plus the power dissipated in the analog-to-digital and digital-to-analog converters in these transceiver chains $\nu B M$. 
We also note that the solution is independent of $\eta$ and $\mu$.

As a side note, we stress that the solution in \eqref{eq:optimal-ratio-PM} and \eqref{eq:optimal-ratio-PM2} is strictly true only if $M$ is allowed to attain a non-integer value. It will in practice only be approximately true.

\section{Single Variable Optimization}

In this section, we show how to optimize the EE with respect to each of the variables $P$, $M$, and $B$ when the other ones are fixed. These results give insights into the solution structure and are the necessary building blocks for developing a joint optimization algorithm in Section~\ref{section:algorithm}. The first result considers optimizing $P$.

\begin{lemma} \label{maximum-EE-P}
The EE in \eqref{eq:EE} for a given $B,M$ is maximized with respect to $P$ by
\begin{equation} \label{eq:optimal-ratio-2}
P = B N_0 \frac{e^{v}-1}{M \beta},
\end{equation}
where
\begin{equation}
v= W \left(\frac{\kappa M \beta(\mu+(D_0+\nu B)M)}{B N_0 e} -\frac{1}{e}\right)+1
\end{equation}

\end{lemma}
\begin{IEEEproof}
The EE with respect to $P$ has a form that can be directly maximized by using \cite[Lem.~3]{bjornson2015a}.
\end{IEEEproof}

By rearranging in \eqref{eq:optimal-ratio-2}, we can once again obtain an expression for $P/B$. However, in this case, $v$ also depends on $B$, so the result is different from Theorem~\ref{maximum-EE-lemma}.

Next, we optimize $B$ when other parameters are fixed. This process can be interpreted as a carrier bandwidth optimization.

\begin{lemma} \label{maximum-EE-B}
The EE in \eqref{eq:EE} is a unimodal function of $B$ (for any fixed $P,M$) that is maximized at a unique $B$. 


\end{lemma}
\begin{IEEEproof}
Since $\mathrm{EE}(B)>0$ for $B>0$ and $\lim_{B\to \infty}\mathrm{EE}(B)=\lim_{M\to 0^+}\mathrm{EE}(B)=0$ there exists a single positive solution $B_\mathrm{opt}$ that minimizes $\mathrm{EE}(B)$. Furthermore, finding $B_\mathrm{opt}$ by setting $\frac{\partial \mathrm{EE}}{\partial B}=0$ leads to the equation
\begin{align}\begin{split}
    &\left(\frac{B N_0}{M P \beta}\left( \kappa \mu +D_0 M  \kappa + P   \right)+\kappa \mu + D_0 M \kappa + P \right) \\
    &\times\log_e\left(1+\frac{M P \beta}{B N_0}\right)= 
     M \kappa \nu B + \kappa \mu + D_0 M \kappa + P\end{split}.
\end{align}
This equation has only one solution since its left-hand side goes to $\infty$ for small $B$ but is always decreasing as $B$ grows and the right-hand side is a positive affine function of $B$. To show that the left-hand side is a monotonically decreasing function, let us take the derivative of it with respect to $B$ and obtain
\begin{align}
  \dfrac{\left({\kappa}{\mu}+D_0M{\kappa}+P\right)\left(BN_0\log_e\left(1+\frac{MP{\beta}}{BN_0}\right)-MP{\beta}\right)}{MP{\beta}B}.  
\end{align}
The above function is always non-positive since $\log_e(1+x)\leq x$ holds for $x>0$, i.e.,
\begin{align}
    B N_0\log_e\left(1+\frac{MP{\beta}}{BN_0}\right)-MP{\beta}\leq 0.
\end{align}
This proves that $\mathrm{EE}(B)$ is a unimodal function of $B$ and that the solution $B_\mathrm{opt}$ can be obtained numerically (e.g., through a bisection search). No closed-form solution exists.
\end{IEEEproof}

\begin{figure}[t!]
	\centering \vspace{-2mm}
	\includegraphics[width=\columnwidth]{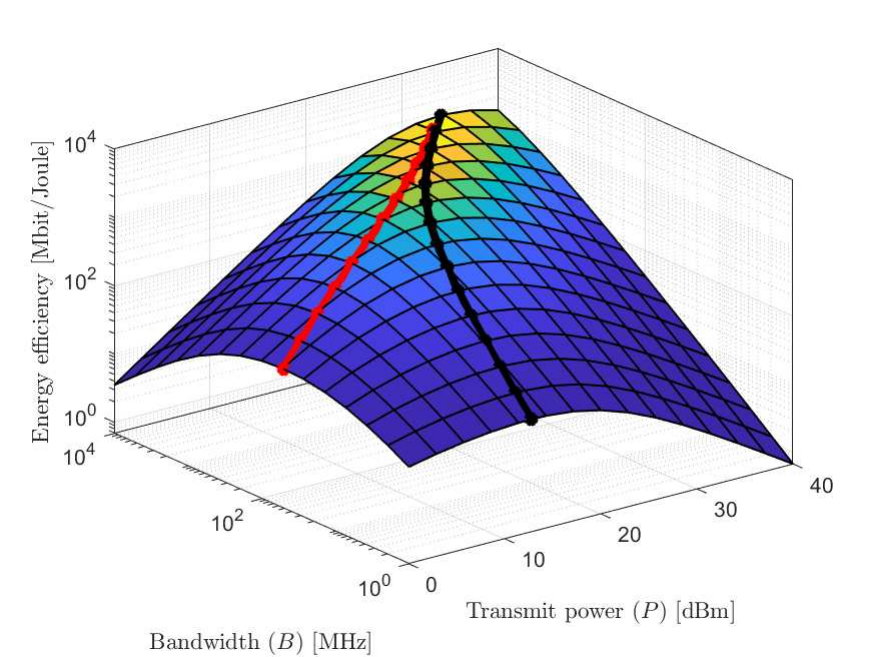}
	\caption{The EE is shown as a function of $P$ and $B$. In black, we have the optimal $P$ for a given $B$ as in Lemma \ref{maximum-EE-P}. In red, we have optimal $B$ for a given $P$ as in Lemma \ref{maximum-EE-B}. $P$ and $B$ converge to the optimal ratio $P/B$ given in Theorem 1. We consider $M=20$.}
	\label{fig:MISOEE} \vspace{-2mm}
\end{figure}

The results of Lemma \ref{maximum-EE-P} and \ref{maximum-EE-B} are illustrated in Fig.~\ref{fig:MISOEE}. In this figure, we consider a fixed number of transmit antennas $M=20 $ and plot the EE for varying values of $P$ and $B$. The optimal $P$ for a given $B,M$ is solved by Lemma \ref{maximum-EE-P} and shown by the black line. The optimal $B$ for a given $P,M$ is solved by Lemma \ref{maximum-EE-B} and shown by the red line. Furthermore, for large values $P$ and $B$, the two lines converge to the optimal ratio as explained in Theorem \ref{maximum-EE-lemma}.

Finally, we optimize $M$ when other parameters are fixed.


\begin{lemma} \label{maximum-EE-M}
The EE in \eqref{eq:EE} for a given $B,P$ is maximized with respect to $M$ by
\begin{equation} \label{eq:optimal-M}
M = B N_0 \frac{e^{w}-1}{P \beta},
\end{equation}
where
\begin{equation}
w= W \left(\frac{P \beta(P/\kappa+\mu)}{B N_0 e(D_0 +\nu B)} -\frac{1}{e}\right)+1
\end{equation}

\end{lemma}
\begin{IEEEproof}
The EE can be directly maximized by using \cite[Lem.~3]{bjornson2015a}. 
\end{IEEEproof}

\begin{figure}[t!]
	\centering \vspace{-2mm}
	\includegraphics[width=\columnwidth]{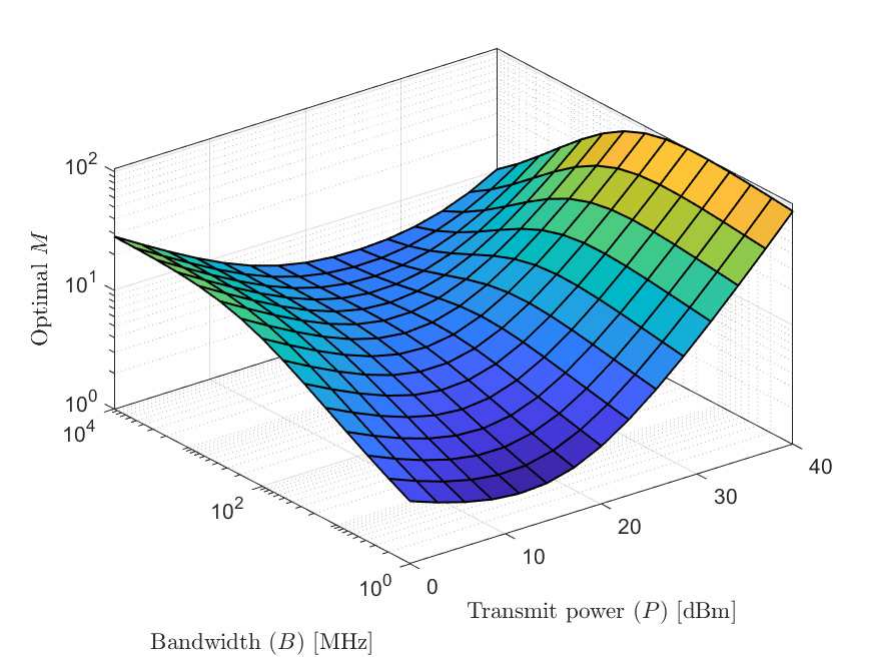}
	\caption{The optimal $M$ as given in Lemma~\ref{maximum-EE-M} as a function of the bandwidth and transmit power. The optimal value is shown on the vertical axis for different $P$ and $B$ values.}
	\label{fig:MisoMstar} \vspace{-2mm}
\end{figure}

In Fig.~\ref{fig:MisoMstar}, we plot the optimal $M$ given by Lemma~\ref{maximum-EE-M} for varying values of $B$ and $P$. We observe that the optimal $M$ attains a wide range of values, depending on $P$ and $B$.

\subsection{Computational efficiency does not affect the solution}

As a corollary to the main results, it is interesting to note that the parameter $\eta$ (i.e., the PC constant proportional to the achieved rate) has no impact on the optimal parameters $(P_\mathrm{opt}, B_\mathrm{opt}, M_\mathrm{opt}) $.

\begin{corollary}
    The optimal solution argument $(P_\mathrm{opt}, B_\mathrm{opt}, M_\mathrm{opt}) $ that maximizes \eqref{eq:EE} is independent of $\eta$.
\end{corollary}

\begin{IEEEproof}
    We define
    \begin{equation} \label{eq:f}
        f=  \frac{B \log_2 \left( 1 + \frac{M P \beta}{B N_0} \right) }{P/\kappa + \mu  + ( D_0+\nu B )M } ,
    \end{equation}
    which is the EE with $\eta=0$.
    The $\mathrm{EE}$ in \eqref{eq:EE} can then be rewritten as
    \begin{equation}
        \mathrm{EE}=\frac{f}{1+\eta f}.
    \end{equation}
    Equating $\mathrm{EE}'=0$ (with respect to any variable) yields
    \begin{equation}
     \frac{f'(1+\eta f)-f(\eta f')}{(1+ \eta f)^2}=0,
    \end{equation}
    which has the only solution $f'=0$. This implies that $\mathrm{EE}$ is maximized precisely when $f$ is maximized, so the optimal parameters are the same.
\end{IEEEproof}

 The consequence is that optimizing $\mathrm{EE}$ in \eqref{eq:EE} can be facilitated by letting $\eta=0$. A similar observation was made in \cite{bjornson2015a} but for a different system model.

\section{Algorithm for optimizing the EE}
\label{section:algorithm}

An algorithm that utilizes our previous results in Lemma 1-3 and which converges to the optimal solution is provided in this section. The goal is to solve the following joint EE maximization problem:

\begin{equation} \label{eq:optimization}
\begin{aligned}
 \underset{P,B,M}{\textrm{maximize}} \quad & \mathrm{EE}\\
\textrm{subject to} \quad & 0 < P \leq P_\mathrm{max},\\
& 0 < B \leq B_\mathrm{max}, \\
 \quad & M\in \{1,\ldots ,M_\mathrm{max}\},
\end{aligned}
\end{equation}
with the $\mathrm{EE}$ defined as in \eqref{eq:EE}.

The following result on whether $P$ or $B$ should be maximized is needed.

\begin{theorem}\label{eitheror}
    The constrained EE maximization problem in \eqref{eq:optimization} is solved at the boundary where $P=P_\mathrm{max}$ or $B=B_\mathrm{max}$.
\end{theorem}

\begin{IEEEproof}
Let us introduce the variables $z$ and $t$ instead of $P$ and $B$ as $z=P/B$, $t=1/P$, and express \eqref{eq:optimization} as 
\begin{equation} \label{eq:varchange}
\begin{aligned}
 \underset{z,t,M}{\textrm{maximize}} \quad & \frac{\frac{1}{z} \log_2 \left( 1 + \frac{M \beta}{N_0} z \right) }{1/\kappa + \mu t + D_0 M t + \frac{\nu M}{z} + \frac{\eta}{z} \log_2 \left( 1 + \frac{M \beta}{ N_0} z \right) }\\
\textrm{subject to} \quad & z \geq 1/P_\mathrm{max},\\
& zt \geq 1/B_\mathrm{max}, \\
 \quad & M\in \{1,\ldots ,M_\mathrm{max}\}.
\end{aligned}
\end{equation}

We see that the problem \eqref{eq:varchange} without the first and second constraints has the optimal solution $t=0$ found outside the feasible region. This disproves the existence of an inner point solution to \eqref{eq:optimization} and proves the theorem.
\end{IEEEproof}



\begin{figure}[t!]
	\centering \vspace{-4mm}
	\includegraphics[width=\columnwidth]{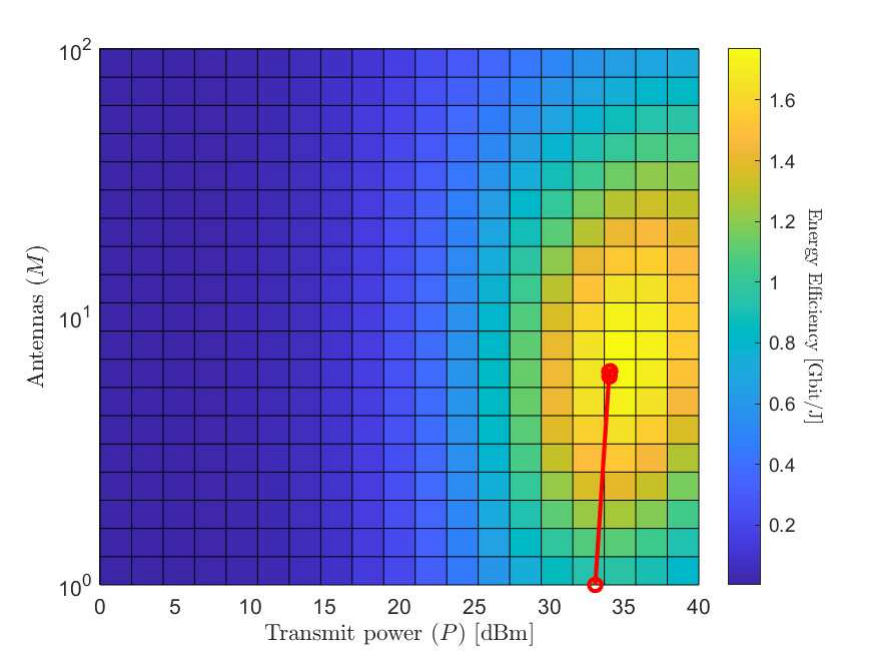}
	\caption{The EE is shown as a function of $P$ and $M$. It converges to the optimum point as detailed in Algorithm 1. Encircled in red we have the algorithm's updates of $P$ and $M$ and the corresponding $\mathrm{EE}$.}
	\label{fig:algo1} \vspace{-2mm}
\end{figure}

We propose Algorithm 1 to solve this problem. Since Theorem \ref{eitheror} establishes that the EE is maximized for either maximum possible transmit power or maximum possible bandwidth, we can invoke Lemma \ref{maximum-EE-P} at maximum bandwidth and compare its solution to Lemma \ref{maximum-EE-B} for maximum power. The solution with the highest EE is used. In the next step (\textit{row 15}), we can use Lemma \ref{maximum-EE-M} to optimize the number of transmit antennas. We can then alternate this optimization until convergence, i.e., again find the best $(B,P)$ as above and again update $M$. If either $P$, $B$, or $M$ becomes too large then it is set to its maximum value. Because the number of transmit antennas should be an integer, in the final step, we consider the two closest possible antenna numbers, compute the respective optimal power and bandwidth,  and select the one achieving the highest EE. The updates of the values of $P$ and $M$ through using Algorithm 1 is provided in Fig.~\ref{fig:algo1}.

 \begin{algorithm}[b!]

 \label{algo1}
 \caption{Joint EE Maximization}
 \begin{algorithmic}[1]
 \renewcommand{\algorithmicrequire}{\textbf{Input:}}
 \renewcommand{\algorithmicensure}{\textbf{Output:}}
 \REQUIRE $\beta, N_0,\mu, \nu,  D_0, \kappa, P_\mathrm{max}, B_\mathrm{max}, M_\mathrm{max}$
 \ENSURE  $P, B, M, \mathrm{EE}$
 \\ \textit{Initialization}: $M=1,\delta=10^{-1}$ bit/J, $i=1$, $\mathrm{EE}_0=0$ bit/J, $\mathrm{EE}_1=1$ bit/J, 
 
 \WHILE {$\mathrm{EE}_{i}-\mathrm{EE}_{i-1}>\delta$} 
 \STATE \textit{use Lemma \ref{maximum-EE-P}}:
   calculate $P$ and $\mathrm{EE}(P,B_\mathrm{max})$
     \IF {($P \geq P_\mathrm{max}$)}
  \STATE $P=P_\mathrm{max}$
  \ENDIF
  
  \STATE \textit{use Lemma \ref{maximum-EE-B}}:
  calculate $B$ and $\mathrm{EE}(P_\mathrm{max},B)$ 
    \IF {($B \geq B_\mathrm{max}$)}
  \STATE $B=B_\mathrm{max}$
  \ENDIF
   
   \IF {$\mathrm{EE}(P_\mathrm{max},B) \geq \mathrm{EE}(P,B_\mathrm{max})$}
  \STATE $P=P_\mathrm{max}$
  \ELSE
  \STATE $B=B_\mathrm{max}$
  \ENDIF
  \STATE \textit{use Lemma \ref{maximum-EE-M}}:  
    update $M$ for these values of $P$ and $B$
        \IF {($M \geq M_\mathrm{max}$)}
  \STATE $M=M_\mathrm{max}$
  \ENDIF
  \STATE update $i=i+1$
  \STATE update $\mathrm{EE_i}(P,B,M)$
  \ENDWHILE
  \STATE \textit{compare the closest integers to $M$} \\
  \STATE update $(P,B)$ for $M=\lfloor M \rfloor $ and $M=\lceil M \rceil $ as in line 2 to 14\\
  \STATE \textit{identify the optimal solution as the one of those two candidate solutions that attain the largest EE}
 \RETURN $P, B, M, \mathrm{EE}_i$ 
 \end{algorithmic} 
 \end{algorithm}

\section{Conclusion}

In this study, we have explored the fundamentals of $\mathrm{EE}$ optimization for wireless communication links. Our investigation has led to several key insights, shedding light on the intricate relationship between power, bandwidth, and the number of transmit antennas. In scenarios without constraints on power or bandwidth, our analysis demonstrates that the ratio of power to bandwidth converges to an optimal value. This implies that an excess of bandwidth does not necessarily lead to improved $\mathrm{EE}$. On the other hand, it implies that we can achieve any data rate simultaneously with the maximum $\mathrm{EE}$, so there is no fundamental tradeoff.


Our results further emphasize that a finite number of transmit antennas offers the highest EE. A novel intriguing discovery is that the total transmit power equals the total transceiver power for the antennas at the optimum, provided that the transmit power does not exceed the maximum limit. To facilitate the application of our findings, we have designed an algorithm that rapidly converges to the optimal solution to the joint EE maximization with respect to transmit power, bandwidth, and antennas. These findings contribute to a deeper understanding of energy efficiency in this field and guide the development of energy-efficient operation of more complex wireless networks.

\bibliographystyle{IEEEtran}

\bibliography{IEEEabrv,Referenser}

\begin{thebibliography}{10}
\providecommand{\url}[1]{#1}
\csname url@samestyle\endcsname
\providecommand{\newblock}{\relax}
\providecommand{\bibinfo}[2]{#2}
\providecommand{\BIBentrySTDinterwordspacing}{\spaceskip=0pt\relax}
\providecommand{\BIBentryALTinterwordstretchfactor}{4}
\providecommand{\BIBentryALTinterwordspacing}{\spaceskip=\fontdimen2\font plus
\BIBentryALTinterwordstretchfactor\fontdimen3\font minus \fontdimen4\font\relax}
\providecommand{\BIBforeignlanguage}[2]{{%
\expandafter\ifx\csname l@#1\endcsname\relax
\typeout{** WARNING: IEEEtran.bst: No hyphenation pattern has been}%
\typeout{** loaded for the language `#1'. Using the pattern for}%
\typeout{** the default language instead.}%
\else
\language=\csname l@#1\endcsname
\fi
#2}}
\providecommand{\BIBdecl}{\relax}
\BIBdecl

\bibitem{ericssonmobilityreport2023}
\BIBentryALTinterwordspacing
Ericsson, ``{Ericsson Mobility Report June 2023},'' 2023. [Online]. Available: \url{https://www.ericsson.com/en/reports-and-papers/mobility-report/reports/}
\BIBentrySTDinterwordspacing

\bibitem{Verdu1990a}
S.~Verd\'u, ``On channel capacity per unit cost,'' \emph{{IEEE} Trans. Inf. Theory}, vol.~36, no.~5, pp. 1019--1030, 1990.

\bibitem{ZapponeNowPublishers2015}
A.~Zappone and E.~Jorswieck, ``Energy efficiency in wireless networks via fractional programming theory,'' \emph{Foundations and Trends in Communications and Information Theory}, vol.~11, no. 3-4, pp. 185--396, 2015.

\bibitem{series2017minimum}
M.~Series, ``Minimum requirements related to technical performance for {IMT-2020} radio interface (s),'' \emph{Report}, vol. 2410, pp. 2410--2017, 2017.

\bibitem{bjornson2017massive}
E.~Bj{\"o}rnson, J.~Hoydis, and L.~Sanguinetti, ``Massive {MIMO} networks: Spectral, energy, and hardware efficiency,'' \emph{Foundations and Trends{\textregistered} in Signal Processing}, vol.~11, no. 3-4, pp. 154--655, 2017.

\bibitem{Rappaport2019a}
T.~S. Rappaport \emph{et~al.}, ``Wireless communications and applications above 100 {GHz}: Opportunities and challenges for {6G} and beyond,'' \emph{IEEE Access}, vol.~7, pp. 78\,729--78\,757, 2019.

\bibitem{RIS-EE}
C.~Huang, A.~Zappone, G.~C. Alexandropoulos, M.~Debbah, and C.~Yuen, ``Reconfigurable intelligent surfaces for energy efficiency in wireless communication,'' \emph{IEEE Transactions on Wireless Communications}, vol.~18, no.~8, pp. 4157--4170, 2019.

\bibitem{lopez2022survey}
D.~L{\'o}pez-P{\'e}rez, A.~De~Domenico, N.~Piovesan, G.~Xinli, H.~Bao, S.~Qitao, and M.~Debbah, ``A survey on {5G} radio access network energy efficiency: Massive {MIMO}, lean carrier design, sleep modes, and machine learning,'' \emph{IEEE Communications Surveys \& Tutorials}, vol.~24, no.~1, pp. 653--697, 2022.

\bibitem{auer2011much}
G.~Auer, V.~Giannini, C.~Desset, I.~Godor, P.~Skillermark, M.~Olsson, M.~A. Imran, D.~Sabella, M.~J. Gonzalez, O.~Blume \emph{et~al.}, ``How much energy is needed to run a wireless network?'' \emph{IEEE Wireless Communications}, vol.~18, no.~5, pp. 40--49, 2011.

\bibitem{Bjornson2016aabb}
E.~Bj{\"o}rnson, L.~Sanguinetti, and M.~Kountouris, ``Deploying dense networks for maximal energy efficiency: {S}mall cells meet massive {MIMO},'' \emph{{IEEE} J. Sel. Areas Commun.}, vol.~34, no.~4, pp. 832--847, 2016.

\bibitem{Zappone2023tradeoff}
A.~Zappone, D.~López-Pérez, A.~De~Domenico, N.~Piovesan, and H.~Bao, ``Rate, power, and energy efficiency trade-offs in massive {MIMO} systems with carrier aggregation,'' \emph{IEEE Transactions on Green Communications and Networking}, vol.~7, no.~3, pp. 1342--1355, 2023.

\bibitem{bjornson2015a}
E.~Bj{\"o}rnson, L.~Sanguinetti, J.~Hoydis, and M.~Debbah, ``Optimal design of energy-efficient multi-user {MIMO} systems: Is massive {MIMO} the answer?'' \emph{IEEE Transactions on Wireless Communications}, vol.~14, no.~6, pp. 3059--3075, 2015.

\bibitem{lopez2021energy}
D.~L{\'o}pez-P{\'e}rez, A.~De~Domenico, N.~Piovesan, X.~Geng, H.~Bao, and M.~Debbah, ``Energy efficiency of multi-carrier massive {MIMO} networks: Massive {MIMO} meets carrier aggregation,'' in \emph{2021 IEEE Global Communications Conference (GLOBECOM)}.\hskip 1em plus 0.5em minus 0.4em\relax IEEE, 2021, pp. 01--07.

\bibitem{piovesan2022machine}
N.~Piovesan, D.~L{\'o}pez-P{\'e}rez, A.~De~Domenico, X.~Geng, H.~Bao, and M.~Debbah, ``Machine learning and analytical power consumption models for {5G} base stations,'' \emph{IEEE Communications Magazine}, vol.~60, no.~10, pp. 56--62, 2022.

\bibitem{piovesan2022power}
N.~Piovesan, D.~Lopez-Perez, A.~De~Domenico, X.~Geng, and H.~Bao, ``Power consumption modeling of {5G} multi-carrier base stations: A machine learning approach,'' \emph{arXiv preprint arXiv:2212.04318}, 2022.

\bibitem{bjornson2018energy}
E.~Bj{\"o}rnson and E.~G. Larsson, ``How energy-efficient can a wireless communication system become?'' in \emph{2018 52nd Asilomar Conference on Signals, Systems, and Computers}.\hskip 1em plus 0.5em minus 0.4em\relax IEEE, 2018, pp. 1252--1256.

\bibitem{Lo1999a}
T.~Lo, ``Maximum ratio transmission,'' \emph{{IEEE} Trans. Commun.}, vol.~47, no.~10, pp. 1458--1461, 1999.

\end{thebibliography}

\end{document}